# Microstructure of Reflection Holographic Grating Inscribed in an Absorptive Azopolymer Film

**Hyunhee Choi**

*Department of Physics, Soongsil University, Seoul 156-743, Korea*

Microstructure of reflection holographic grating fabricated via a photo-isomerization process in an absorptive azopolymer film is analyzed. A surface relief formation takes place on the film surface even in the reflection holographic configuration. The polarization-dependent diffraction efficiency and the polarization analysis reveal that the polarization grating structure inside the film strongly depends on the amount of optical absorption experienced by the two writing beams. Theoretical analysis shows that the reflection polarization grating, while mimicking a cholesteric liquid crystal structure, is composed of elliptic polarizations with the ellipticity going through a periodic modulation.

# I. INTRODUCTION

The ability to control optical birefringence through a photo-isomerization process renders azobenzene-containing polymer an important optical material class for use in the optical memory[1,2], imaging of the electromagnetic field distributions[3,4], photoresists[5,6], and high efficiency holographic recording[7]. Since the report of the surface relief grating(SRG) fabrication by a holographic writing of two interfering beams[8], the characteristics of the resultant diffraction grating have been analyzed. In most works employing the photo-isomerization process, a transmission holographic configuration has been adopted, because the optical absorption by the azobenzene moiety in the polymer does not allow a deep penetration of the hologram into the polymeric film. However, other kinds of optical materials such as liquid crystalline polymer[9], azobenzene-containing gelatine[10] as well as a far off-resonance writing at a thick polymeric film [11] have been adopted to implement a reflection holographic configuration in forming a diffraction grating, and the fabrication of both intensity- and polarization-modulation gratings has been successfully demonstrated. While the grating vector of the formed diffraction grating is parallel to the film surface in the transmission configuration, it is perpendicular to the film surface in the reflection configuration, which makes the diffraction grating fabricated by the reflection configuration find applications in new architecture photonic devices. However, broader application can be expected if it can form the grating on the surface and inside simultaneously.

In this work, a slanted reflection configuration is adopted to write a reflection hologram in the optically absorptive regime of azo-benzene moiety to explore how the photo-isomerization process affects the diffraction grating formation both in the bulk and on the surface.
We report two important experimental findings in the slanted reflection gratings fabricated by two counter-propagating co-circularly polarized beams. First, SRG formation takes place on the film surface even in a reflective holographic beam configuration. Second, owing to the differences in the optical absorption experienced by two counter-propagating beams at a given interfering plane, the ellipticity of polarization of the reflection hologram goes through a periodic modulation along the film depth, giving rise to an ellipticity dependence of the diffraction efficiency.

## II. EXPERIMENTS AND DISCUSSION

A commercially available, azobenzene-pendant side-chain polymer, poly-(disperse-red 1-methacrylate-co-methyl-methacrylate), was employed in this experiment, whose glass transition temperature $T_g$ is 131°C. By spin coating a 10 wt. % tetrahydrofurane solution of the polymer onto a glass substrate at 2500 rpm and drying for 1h at 100 °C, a 1 $\mu$m thick polymeric film was readily obtained with the absorption peak at 490nm and the optical density ~1.5 at λ=514.5 nm. For writing, two counter-propagating beams of 514.5 nm line from an Ar$^+$ laser were used with the light intensity of 200 mW/cm$^2$ on the film. The polarization states of writing beams are controlled by a combination of the half- and quarter-wave plates. The cross angle of two beams is 1~2° corresponding to an interference modulation with the periodicity Λ=257 nm (λ/2). The polymer film is slanted with the grating slant angle $\theta_s$=15° with respect to the grating vector. In Fig. 1(a), a photograph of the reflection holographic setup is shown. In the slanted reflection holographic configuration, one can observe the reflected and transmitted diffraction orders, **±1$_R$**, **±1$_T$**, and **0$_T$**, and **0$_R$**, as seen in Fig. 1(b). The diffraction orders are monitored by an S-polarized 632.8nm He-Ne laser at an intensity < 0.1 mW, incident from substrate in parallel to the grating vector, chopped at 1 kHz. Two identical photodiodes in conjunction with two lock-in amplifiers are used to simultaneously monitor both transmitted and reflected orders.

Once the formation of a diffraction grating is identified by the appearance of diffraction orders, typical diffraction efficiency (DE) in the order of 1 %, the surface morphology of the fabricated grating polymer film was studied by the atomic force microscope (AFM).

Fig.2 (a) shows schematics illustration of slanted reflection grating formation. AFM images of the SRGs, fabricated with the recording time of 20 minutes at the irradiation intensity of 200 mW/cm$^2$, were obtained by a commercial AFM (PSIA, XE-100). Fig. 2(b) shows AFM images of the SRGs fabricated in counter propagating co-circular (C/C) polarization configuration. In the current holographic geometry where two writing beams are counter-propagating along a straight line, a polarization modulation, not an intensity modulation, takes place on the surface of the slanted film as shown in Fig. 2(c). We find that C/C provides a surface relief much more distinct than that formed by S/P polarization configuration, similar to that observed in the surface reliefs fabricated in a transmission holographic configuration [12].

Now in order to clarify the polarization optical structure underneath the SRG, we prepared two azopolymer films possessing different optical absorptions, at the writing beam of Ar$^+$ laser 514.5 nm line, namely, 7 wt. % and 10 wt. % solutions are employed to spin-coat a low-absorbing (LA) and a high-absorbing (HA) films. Both LA and HA films are illuminated for 1 minute to write a slanted reflection hologram in the C/C polarization configuration. The diffraction efficiencies (DEs) of $\pm 1_R$, $\pm 1_T$, and $0_T$, and $0_R$ are measured as a function of the ellipticity $\varepsilon$, of the probe beam of He-Ne laser 633nm line. At the same time, the polarization analysis of the diffraction orders is carried out. See Fig. 1(b) for the designation of diffracted orders. Fig. 3(a) &(c) show $\varepsilon$ dependence of DE, and Fig. 3(b) &(d) show the polarization changes of $+1_R$ for S-, P-, LCP, and RCP polarizations, with (a) &(b) for LA and (c) &(d) for HA films. For both LA and HA films, $+1_R$ order has the highest DE among the diffraction orders, with the peak magnitudes in a similar size. However, the ellipticity $\varepsilon$ dependence is quite different. Namely, DE of $+1_R$ order from LA film has the maximum peak between RCP and P-polarization, while the maximum peak occurs between S- and RCP polarization for film sample. The polarization analysis of $+1_R$ order shows that the LCP preserves the polarization, while the RCP converts to linear polarization for both samples. Linear S- and P-polarizations, however, go through a rotation of polarization plane by ~30° and ~10° for LA and HA films, respectively. On the other hand, **-1$_R$** orders have DEs lower than those of **+1$_R$** orders, ~50% and ~10% for LA and HA films, respectively. As far as ε dependence is concerned, one striking feature is that the maximum peak of DE occurs at S-polarization for both LA and HA films, which suggests that **-1$_R$** order is more affected by the surface relief, and less affected by the polarization grating structure.

The **+1$_R$** order is from the reflection normal to the reflective grating and, accordingly, contains information on the polarization grating structure inside the film bulk. From this, we suppose that the optical absorptions experienced by the two counter-propagating C/C writing beams play an important role in determining the polarization ellipsoid structure in the polarization grating. Let's look at how an optical absorption affects the polarization structure of a reflective holographic grating. When two counter-propagating circularly polarized beams, $\boldsymbol{E_r}^+ = \xi_+(\hat{\boldsymbol{x}} + i\hat{\boldsymbol{y}})\exp\{i(\omega t - kz)\}$ and $\boldsymbol{E_r}^- = \xi_-(\hat{\boldsymbol{x}} - i\hat{\boldsymbol{y}})\exp\{i(\omega t + kz)\}$, are employed for holographic recording, a superimposed wave at the depth $z$ takes the form,

$$\boldsymbol{E}(z,t) = \boldsymbol{E_r}^+ + \boldsymbol{E_r}^- = \xi_+(\hat{\boldsymbol{x}} + i\hat{\boldsymbol{y}})\exp\{i(\omega t - kz)\} + \xi_-(\hat{\boldsymbol{x}} - i\hat{\boldsymbol{y}})\exp\{i(\omega t + kz)\} \quad (1)$$

with $\xi_+ = E_0 \exp\{-\alpha(d-z)\}$, $\xi_- = E_0 \exp(-\alpha z)$, $d$ thickness of the azo-polymer film, $\alpha$ the optical absorption coefficient, and $E_0$ the initial amplitude of the two writing beams. In the absence of an optical absorption ($\alpha = 0$), $\boldsymbol{E}(z,t)$ reduces to a simple form

$$\boldsymbol{E}(z,t) = 2E_0\{\hat{\boldsymbol{x}}(\cos\omega t \cos kz + i\sin\omega t \cos kz) + i\hat{\boldsymbol{y}}(\cos\omega t \cos kz + i\sin\omega t \cos kz)\} \quad (2)$$

The polarization state that corresponds to a filled helicoidal standing wave structure is shown in Fig.4(b). The ensuing photo-isomerization process replicates the filled helicoidal standing wave structure, giving rise to a polarization grating inside the film, which is the very cholesteric liquid crystal (CLC) structure with a periodicity $\Lambda_{\text{volume}} = \lambda/2$ along the film depth [10, 13].

When an optical absorption from the azopolymer film is significant, however, the polarization state of the resulting reflection polarization grating is rather complicated. At the middle point of the film depth ($z = d/2$), the amplitudes of two writing beams are the same after getting attenuated by the same amount, which results in a linear polarization state. In the interfering planes located at points other than the middle, the unequal amplitudes of two writing beams give rise to an elliptic polarization state. Fig.4(b) shows the theoretical simulation of polarization states, where we note that the ellipticity $\varepsilon$ of polarization goes through a periodic modulation along the film depth as shown in Fig. 4(a). Importantly, $\varepsilon$ at the front and end surfaces of the film, that is, the amount of deviation from the linear polarization, is determined by the amount of optical absorption present and experienced by two writing beams. In the film with a weaker absorption, a smaller deviation from the linear polarization takes place, namely, the polarization state is less elliptical, implying that the overall polarization grating is less distorted from the CLC structure.

Very high DEs of RCP for both LA and HA films, as can be seen in Fig.3(a)&(c), suggest that the reflection grating is mainly in a structure of right-handed CLC, though not perfect as indicated by the non-preservation of the polarization state. In case of LCP, the DE is very low and the polarization state is preserved, which also conforms to the signature of right-handed CLC structure. The fact that DE of RCP for LA film is higher than that for HA film is consistent with the fact that LA film is more CLC-like in the structure than HA film [6]. Furthermore, the observed rotation of polarization planes of linear S- and P-polarizations in **+1$_R$** also implies the CLC structure of the reflection grating. A larger rotation angle for LA film, compared to HA film, tells us that LA film is structured in a way more similar to CLC than HA film, which can be understood from the theoretical analysis discussed above.

## III. CONCLUSION

In conclusion, the microstructure of a reflection holographic grating fabricated via a photo-isomerization process is analyzed. A surface relief formation is observed to take places on the film surface, and optical analyses show that the polarization grating structure inside the film strongly depends on the amount of optical absorption experienced by the two writing beams. It is found that the reflection polarization grating, while mimicking a cholesteric liquid crystal structure, is composed of elliptic polarizations with the ellipticity going through a periodic modulation.

## ACKNOWLEDGEMENT

This research was supported by Basic Science Research Program through the National Research Foundation of Korea (NRF) funded by the Ministry of Science, ICT and Future Planning (No. NRF-2012R1A1A3010975).

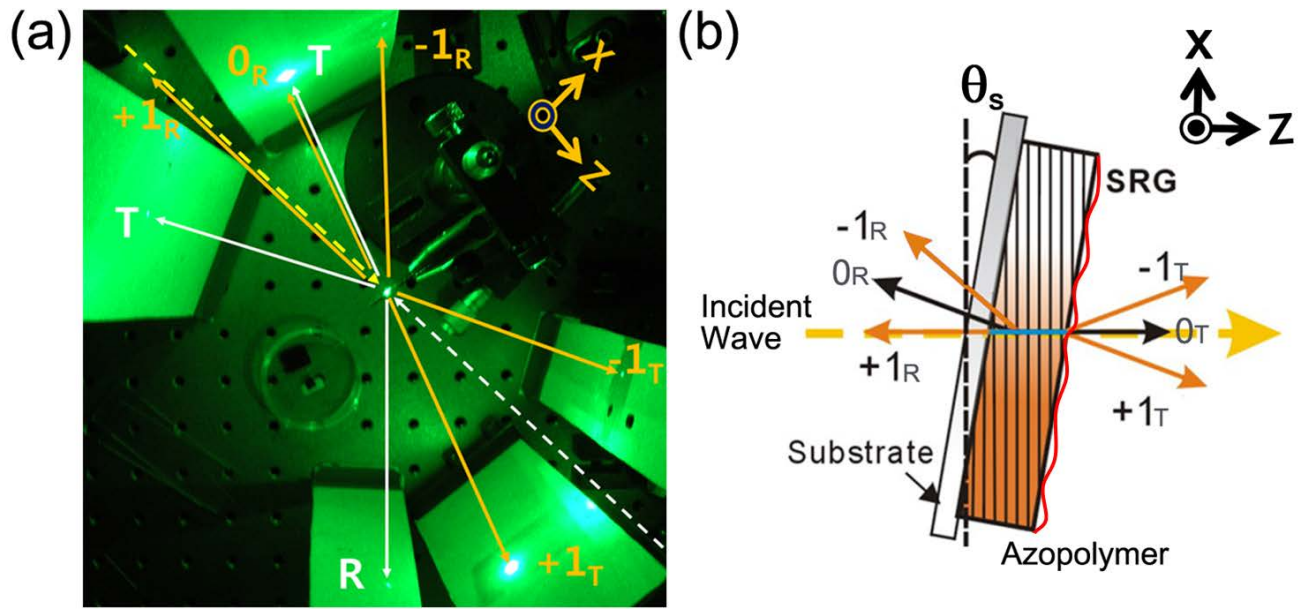

Fig. 1. (a) Photograph of reflection holographic setup. T and R stand for the transmitted and reflected light, respectively. (b) Schematic diagram of the reflection hologram formed on the azopolymer film. SRG stands for surface relief grating, and **+1R, -1R, +1T,** and **-1T** correspond to each diffraction orders.

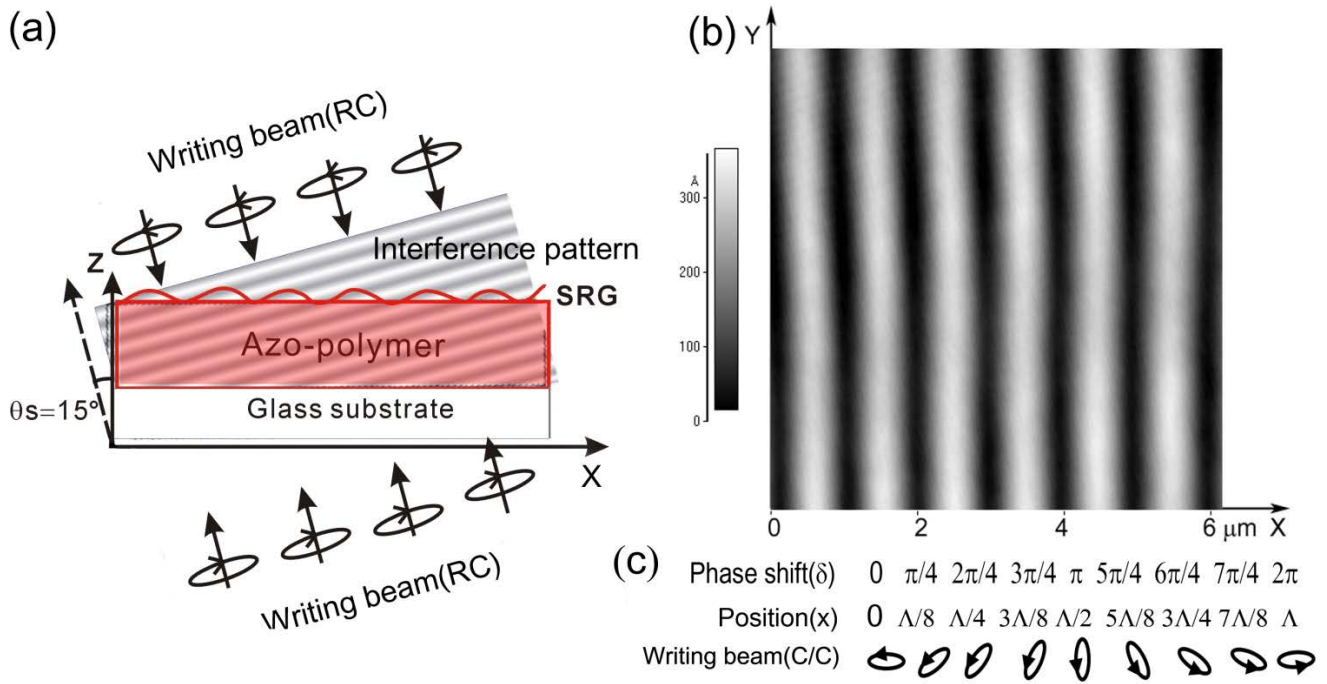

Fig. 2. (a) Schematic illustration of slanted reflection grating formation. (b) AFM images of SRGs. (c) Types of polarization modulation in recording with two waves with counter propagating RCP and RCP on the surface.

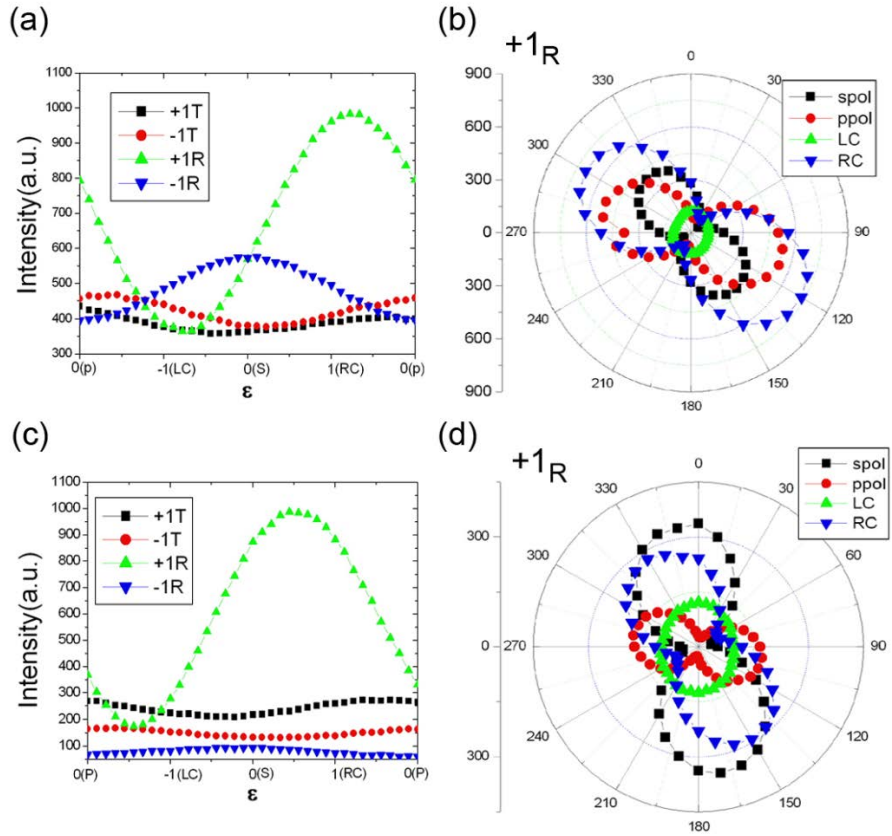

Fig. 3. (a)&(c) show the ellipticity dependence of DE, and (b)&(d) show the polarization changes of +**1R** for S-, P-, LCP, and RCP polarizations, with (a)&(b) for LA and (c)&(d) for HA films.

(a)

Phase shift(δ)    0  π/4  2π/4  3π/4  π  5π/4  6π/4  7π/4  2π
Position(z)       0  Λ/8  Λ/4  3Λ/8  Λ/2  5Λ/8  3Λ/4  7Λ/8  Λ
Writing beam(C/C)

(b)

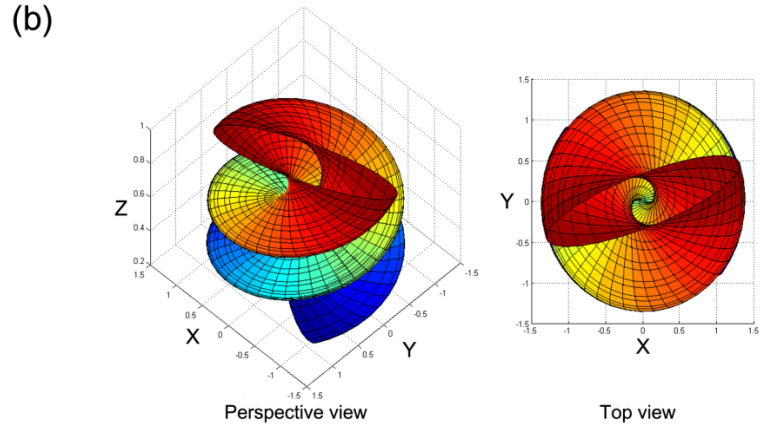

Perspective view　　　　　　　Top view

Fig.4. (a) Types of polarization modulation and (b) the helicoidal standing wave interference resulting from the of counter propagating RCP waves when an optical absorption is present.